\documentclass{article} % For LaTeX2e
\usepackage{iclr2027_conference,times}

\usepackage{amsmath,amsfonts,bm}

\def\eqref#1{equation~\ref{#1}}
\def\1{\bm{1}}

\DeclareMathAlphabet{\mathsfit}{\encodingdefault}{\sfdefault}{m}{sl}
\SetMathAlphabet{\mathsfit}{bold}{\encodingdefault}{\sfdefault}{bx}{n}

\usepackage{url}
\usepackage{mdframed}
\usepackage{multirow}
\usepackage{caption}
\usepackage{makecell}
\usepackage{graphicx}
\usepackage[table]{xcolor}

\usepackage{subfigure}
\usepackage[utf8]{inputenc} % allow utf-8 input
\usepackage[T1]{fontenc}    % use 8-bit T1 fonts
\usepackage{url}            % simple URL typesetting
\usepackage{booktabs}       % professional-quality tables
\usepackage{amsfonts}       % blackboard math symbols
\usepackage{nicefrac}       % compact symbols for 1/2, etc.
\usepackage{microtype}      % microtypography
\usepackage{xcolor}         % colors

\usepackage{pifont}
\usepackage{wrapfig}
\usepackage{algorithm}
\usepackage{algorithmic}
\usepackage{amsmath}
\usepackage{enumitem}
\usepackage{fontawesome7}
\usepackage{amsthm}
\newtheorem{definition}{Definition}

\usepackage[most]{tcolorbox}

\usepackage[
    colorlinks=true,
    citecolor=cyan,
    linkcolor=cyan,
    urlcolor=cyan
]{hyperref}

\title{Concealing LLM-Based Multi-Agent\\Topology via Phantom Structure Injection}

\author{Longzhu He\\Nanyang Technological\\University
\And
Zelang Wen\\Beijing University of Posts\\and Telecommunications
\And
Xinfeng Li\\The Hong Kong\\Polytechnic University
\AND
Sen Su\\Beijing University of Posts and Telecommunications
\And
XiaoFeng Wang\\Nanyang Technological University
}

\iclrfinalcopy 
\begin{document}

\maketitle

\begin{abstract}
Driven by the rapid advancement of large language models (LLMs), LLM-based multi-agent systems (MAS) have emerged as a powerful paradigm for collaborative reasoning over complex tasks. A key design element of MAS is the communication topology, which governs information flow among agents and often encodes proprietary knowledge about the system architecture. However, recent work has shown that such topologies can be inferred even in black-box settings by exploiting semantic dependencies in observable reasoning traces, posing significant risks of intellectual property leakage and exposure of system vulnerabilities. To address this threat, we propose \textsc{Mirage}, a topology-concealment framework that preserves the genuine communication topology for task execution while shaping adversary-facing semantic evidence toward a carefully constructed phantom topology. Specifically, \textsc{Mirage} operates in three stages: \ding{172} phantom topology synthesis, \ding{173} semantic edge realization, and \ding{174} protected MAS execution. It constructs a phantom topology structurally distinct from the genuine one, materializes phantom edges as plausible semantic dependencies, and suppresses source-specific cues that could reveal genuine edges absent from the phantom topology. Extensive experiments across three topology optimization frameworks and four benchmark datasets demonstrate that \textsc{Mirage} substantially reduces the effectiveness of topology inference attacks while largely preserving the task utility of the protected MAS.
\end{abstract}

\vspace{-1em}
\section{Introduction}
\vspace{-0.5em}

The rapid advancement of large language models (LLMs)~\citep{naveed2025comprehensive} has given rise to LLM-based multi-agent systems (MAS)~\citep{li2024survey,he2025llm,shao2026multi,DBLP:conf/www/DuHHZLGKY26}, which harness collaborative reasoning and division of labor among multiple agents to achieve remarkable performance on complex, multi-step tasks. Compared with single-agent approaches~\citep{yang2024swe,shinn2023reflexion,zhao2024expel}, MAS decompose complex analytical tasks into specialized subtasks handled by agents with distinct expertise, demonstrating strong capabilities in data-intensive scenarios such as data analysis~\citep{bai2025insight}, mathematical reasoning~\citep{lei2024macm}, and code generation~\citep{islam2024mapcoder}. These advantages have established MAS as an increasingly important technical paradigm for frontier applications, including data discovery~\citep{xiao2026unified}, web data collection~\citep{ma2026autodata}, and scientific discovery~\citep{ghareeb2026multi}.

The effectiveness of MAS largely stems from their carefully designed \textit{communication topology}, which defines the directed connectivity structure governing information flow among agents. By determining how agents interact and how information propagates through the system, the topology directly shapes the system's reasoning process and task performance. Recent studies~\citep{zhang2025g,DBLP:conf/ecai/LiZLW25,li2026assemble} have therefore explored topology optimization methods that automatically construct effective communication structures for specific tasks. These carefully optimized topologies can thus encode substantial design expertise and constitute valuable intellectual property (IP).

\begin{figure}[t]
  \centering
\includegraphics[width=\linewidth]{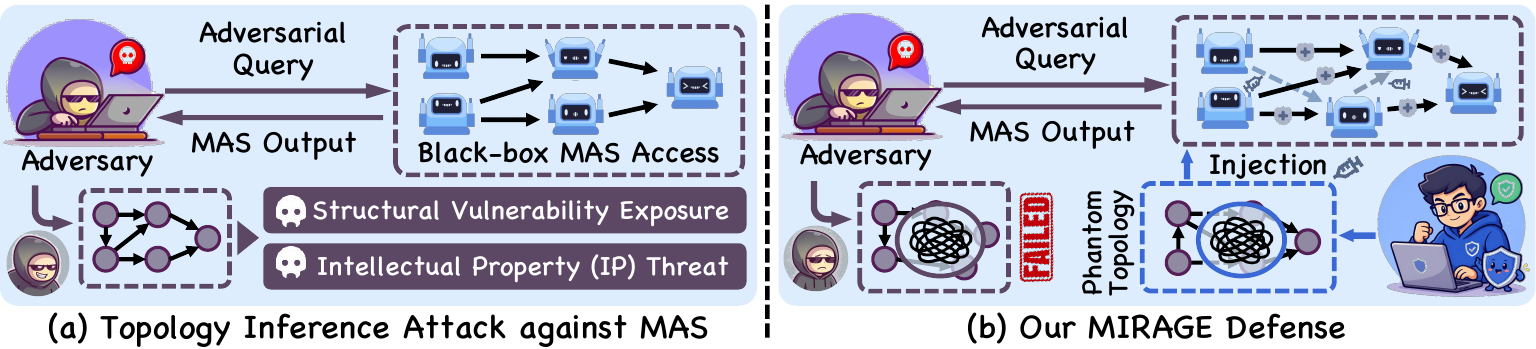}
\vspace{-0.5em}
\caption{\textbf{Comparison of topology inference attack and \textsc{Mirage} defense.} \textbf{(a)} The topology inference attack against LLM-based MAS exploits semantic dependencies in observable reasoning traces to infer the genuine communication topology $\mathcal{G}$. \textbf{(b)} \textsc{Mirage} constructs a phantom topology $\mathcal{G}'$ to mislead topology inference while preserving $\mathcal{G}$ for actual task execution. This design decouples topology exposure from task execution, effectively concealing the genuine communication topology.
}
\vspace{-1em}
\label{fig1}
\end{figure}

However, as MAS are increasingly deployed in real-world applications, their communication topology introduces a largely overlooked security concern. Existing research has primarily focused on attacks targeting communication content, such as malicious prompt propagation~\citep{lee2025prompt} and communication manipulation~\citep{he2025red,ju2026flooding}, while the confidentiality of the communication topology itself has received limited attention. Recently, CIA~\citep{wu2026cia}, as illustrated in Fig.~\ref{fig1}(a), revealed that an adversary can infer the internal communication topology of an MAS solely through black-box interactions. Its key insight is that directly connected agent pairs tend to exhibit stronger semantic dependencies in their observable reasoning outputs, which can be exploited to reconstruct the underlying topology. Such topology leakage not only exposes proprietary system architectures but also reveals critical agents and communication pathways, enabling adversaries to launch targeted attacks that can compromise the security of the MAS.

In light of this threat, an important research question arises: \textit{can we effectively protect the communication topology of an MAS from inference attacks without compromising its task performance?} This problem is challenging because an effective defense method must reconcile two seemingly conflicting objectives: disrupting the semantic evidence that reveals the true topology while preserving the information flow required for task execution. Existing MAS defense studies primarily address content-level security~\citep{zhang2024psysafe,miao2025blindguard,raza2026trism,zhou2026guardian}, leaving systematic protection against communication topology inference attacks largely unexplored.

To address this gap, we propose \textsc{Mirage}, a topology-concealment framework that separates the topology governing task execution from the topology exposed to adversaries. As shown in Fig.~\ref{fig1}(b), \textsc{Mirage} proactively constructs a phantom topology $\mathcal{G}'$ that is structurally distinct from the genuine topology $\mathcal{G}$. Specifically, \textsc{Mirage} consists of three stages:  \ding{172} \textit{phantom topology synthesis}, \ding{173} \textit{semantic edge realization}, and \ding{174} \textit{protected MAS execution}. The first stage synthesizes the phantom topology $\mathcal{G}'$ under structural constraints to ensure sufficient divergence from $\mathcal{G}$. The second stage materializes phantom edges as plausible, task-relevant semantic dependencies while suppressing the source-specific semantic signatures of genuine edges absent from $\mathcal{G}'$. The final stage executes tasks strictly according to $\mathcal{G}$ while generating adversary-facing traces through a separate exposure view shaped toward the phantom topology $\mathcal{G}'$. This design preserves the original communication structure for task execution while obscuring the genuine topology from inference attacks~\citep{wu2026cia}. We evaluate \textsc{Mirage} on four representative datasets across three domains: general reasoning (\texttt{MMLU}~\citep{DBLP:conf/iclr/HendrycksBBZMSS21}), mathematical reasoning (\texttt{GSM8K}~\citep{cobbe2021training} and \texttt{SVAMP}~\citep{patel2021nlp}), and code generation (\texttt{HumanEval}~\citep{chen2021evaluating}). Experimental results show that \textsc{Mirage} substantially reduces the effectiveness of topology inference attacks while largely preserving the task utility of the protected MAS. The main contributions of this paper are summarized as follows:

\begin{itemize}[leftmargin=*, itemindent=0em] 
\renewcommand{\labelitemi}{$\diamond$} 
\item \textit{Important Problem.} This paper investigates the communication topology confidentiality problem in LLM-based MAS, highlighting the security and intellectual property risks posed by topology inference attacks and addressing an underexplored security dimension of multi-agent systems.
\item \textit{Novel Defense.} The proposed \textsc{Mirage} separates task execution from adversary exposure through phantom topology synthesis, semantic edge realization, and dual-view protected execution.
\item \textit{Extensive Evaluation.} Extensive experiments cover three representative MAS topology optimization frameworks and four benchmark datasets across diverse reasoning and code generation tasks: \texttt{MMLU}, \texttt{GSM8K}, \texttt{SVAMP}, and \texttt{HumanEval}. The results demonstrate that \textsc{Mirage} substantially reduces topology inference effectiveness while largely preserving the task utility.
\end{itemize}

\section{Preliminaries}\label{sec2}

\vspace{-0.5em}

This section introduces the essential background on LLM-based multi-agent systems and topology inference attacks, followed by the problem definition. Important notations are summarized in App.~A.

\textbf{LLM-based MAS.} An LLM-based MAS is formalized as $\mathcal{S} = (\mathcal{P}, \mathcal{G})$, where $\mathcal{P} = \{p_i\}_{i=1}^n$ denotes the set of agent profiles, with each $p_i$ specifying the corresponding agent's system prompt, callable tools, and other configuration details. The communication topology is represented as a directed acyclic graph (DAG) $\mathcal{G} = (\mathcal{A}, \mathcal{E})$, where $\mathcal{A} = \{a_i\}_{i=1}^n$ is the set of agents and $\mathcal{E}$ is the set of directed communication edges. An edge $(a_j, a_i) \in \mathcal{E}$ indicates that the output of agent $a_j$ is passed to agent $a_i$ as part of its input. By governing how information is propagated and aggregated across agents, the communication topology plays a central role in MAS performance~\citep{zhang2025g,DBLP:conf/ecai/LiZLW25,li2026assemble}. In practical deployments, a carefully optimized topology may therefore constitute valuable intellectual property for system developers. Given a task query $q$, the output of agent $a_i$ is defined as $r_i = \text{LLM}(p_i, q, \mathcal{O}_i)$, where $\mathcal{O}_i = \{ r_j \mid (a_j, a_i) \in \mathcal{E} \}$ denotes the set of outputs received from its predecessor agents. The final output of the system $\mathcal{S}$ is produced by the designated decision agent $a_n$: \begin{equation} r_n = \mathcal{S}(q) =\text{LLM}(p_n, q, \mathcal{O}_n). \end{equation}

\vspace{-0.5em}

\textbf{Topology Inference Attack.} The topology inference attack aims to reconstruct the hidden communication topology $\mathcal{G}$ of an MAS $\mathcal{S}$ under black-box access, relying only on adversarial queries and the corresponding system responses. A representative attack is \textit{CIA}~\citep{wu2026cia}, which first crafts specially designed adversarial queries to induce the MAS to expose the intermediate reasoning outputs of its agents, and then exploits the semantic dependencies among these outputs to infer the underlying communication edges. More details of the CIA attack are provided in App.~B.

\textbf{Problem Definition.}
Given an LLM-based MAS $\mathcal{S}=(\mathcal{P},\mathcal{G})$ with a private topology $\mathcal{G}$, we consider a black-box adversary  that attempts to reconstruct $\mathcal{G}$ via topology inference. This work aims to protect $\mathcal{G}$ from topology inference while preserving its use for task execution. Specifically, we seek to construct a protected system $\widetilde{\mathcal{S}}$ that reshapes adversary-facing evidence such that the inferred topology $\widehat{\mathcal{G}}$ deviates from the genuine topology $\mathcal{G}$. Formally, the defense aims to maximize the discrepancy between the inferred and genuine topologies, subject to bounded degradation in task utility:
\begin{equation}
\max_{\widetilde{\mathcal{S}}}\;
\Delta_{\text{topo}}(\widehat{\mathcal{G}},\mathcal{G})
\quad
\text{s.t.}\quad
U(\widetilde{\mathcal{S}})\geq U(\mathcal{S})-\epsilon,
\end{equation}
where $\widehat{\mathcal{G}}$ denotes the topology inferred by the adversary, $\Delta_{\text{topo}}(\cdot,\cdot)$ measures the discrepancy between two topologies, $U(\cdot)$ denotes task utility, and $\epsilon$ specifies the maximum allowable utility degradation.

\vspace{-0.5em}
\section{Threat Model}\label{sec3}

\vspace{-0.5em}

This section defines the threat model by specifying the adversary's objectives and capabilities.

\textbf{Adversary's Objectives.} The adversary aims to reconstruct the true communication topology $\mathcal{G}$ of the target multi-agent system $\mathcal{S}$. Successful topology inference may lead to two major security risks:

\vspace{-0.5em}
\begin{itemize}[leftmargin=*, itemindent=0em] 
\renewcommand{\labelitemi}{$\diamond$} 
\item \textit{Structural Vulnerability Exposure.} Once the communication topology $\mathcal{G}$ is revealed, the adversary can identify critical agents, such as those with high in-degree or betweenness centrality, and important communication pathways. Such structural knowledge can facilitate targeted attacks, including jailbreaking~\citep{DBLP:conf/icml/GuZPDL00L24,shahroz2025agents}, prompt injection~\citep{lee2025prompt,he2025red,arif2026conjunctive}, and other attacks~\citep{kavathekar2026tamas} against the MAS.
\item \textit{Intellectual Property (IP) Threat.} A carefully optimized communication topology $\mathcal{G}$ often embodies substantial computational investment and expert design knowledge, making it a valuable proprietary asset of system developers~\citep{li2026assemble,zhang2025g,DBLP:conf/ecai/LiZLW25}. Leakage of the topology may expose internal architectural design information, enable unauthorized replication of the system, and undermine the competitive advantage of the system owner.
\end{itemize}
\vspace{-0.5em}

\textbf{Adversary's Capabilities.} The adversary operates under a strict black-box setting. Specifically, the adversary can submit queries of its choice to the target system $\mathcal{S}$ and observe the corresponding final response $\mathcal{S}(q)$. The adversary has no direct access to internal information, including agent profiles $\mathcal{P}=\{p_i\}_{i=1}^{n}$, system prompts, intermediate communication messages, or the true communication topology $\mathcal{G}$. Nevertheless, the adversary may craft adversarial queries that induce the system to expose information related to intermediate agent outputs in its final response, which can subsequently be exploited for topology inference. Furthermore, the adversary cannot modify agent configurations, communication edges, or any other internal component of $\mathcal{S}$ during the attack process.

\begin{figure}[t]
  \centering
\includegraphics[width=\linewidth]{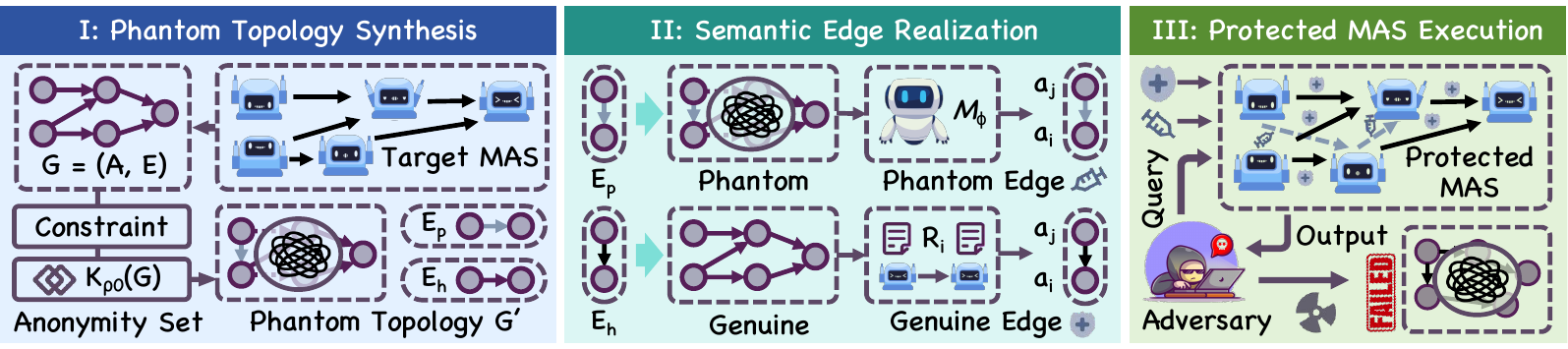}
\vspace{-1.5em}
  \caption{\textbf{Overview of \textsc{Mirage}.} \textsc{Mirage} consists of three stages: \ding{172} phantom topology synthesis, \ding{173} semantic edge realization, and \ding{174} protected MAS execution. Stage I constructs a phantom topology $\mathcal{G}'$ that deviates from the genuine topology $\mathcal{G}$. Stage II realizes $\mathcal{G}'$ by shaping adversary-facing semantic dependencies. Stage III preserves $\mathcal{G}$ for task execution while exposing dependency evidence aligned with $\mathcal{G}'$, thereby concealing the genuine topology with minimal impact on task utility.}
  \vspace{-1em}
\label{fig:2}
\end{figure}

\vspace{-2em}
\section{Methodology}\label{sec4}
\vspace{-0.5em}
To defend against topology inference attacks, we propose \textsc{Mirage}, a topology-concealment framework that decouples the communication structure used for actual task execution from that exposed to the adversary. \textsc{Mirage} consists of three stages: \ding{172} \textit{Phantom Topology Synthesis}, \ding{173} \textit{Semantic Edge Realization}, and \ding{174} \textit{Protected MAS Execution}. The overall workflow is illustrated in Fig.~\ref{fig:2} and Alg.~\ref{alg:mirage}.

\vspace{-0.5em}
\subsection{Phantom Topology Synthesis}\label{sec4.1}
\vspace{-0.5em}
At this stage, \textsc{Mirage} carefully constructs a phantom topology $\mathcal{G}'=(\mathcal{A},\mathcal{E}')$ to guide adversary-facing evidence away from the genuine topology $\mathcal{G}$. We preserve the original agent set $\mathcal{A}$ and perturb only communication edges, avoiding additional agents, roles, or execution traces. This edge-level design minimizes changes to the underlying MAS while more directly targeting the confidential communication structure. An effective phantom topology should sufficiently differ from $\mathcal{G}$ to hinder topology recovery while remaining structurally feasible as a legitimate MAS topology. Accordingly, we formulate its construction as a constrained topology reconstruction problem.

Specifically, let $\pi(\cdot)$ denote the execution order induced by $\mathcal{G}$. We first define the admissible edge space as $\mathcal{U}_{\pi}=\{(a_j,a_i)\mid\pi(a_j)<\pi(a_i)\}$, ensuring that candidate edges respect the execution order and do not introduce cycles. Accordingly, the feasible topology set is defined as
\begin{equation} 
\Omega(\mathcal{G})=\{\mathcal{H}=(\mathcal{A},\mathcal{E}_{\mathcal H})\mid\mathcal{E}_{\mathcal H}\subseteq\mathcal{U}_{\pi},\ \forall a_i\neq a_n,\ a_i\leadsto_{\mathcal H}a_n\}, 
\end{equation} 
where $a_i\leadsto_{\mathcal H}a_n$ indicates that a directed path exists from agent $a_i$ to the decision agent $a_n$ under $\mathcal{H}$. The first constraint ensures that the synthesized topology remains acyclic, while the second prevents isolated agents or branches that cannot contribute to the final decision. To quantify the structural deviation of a candidate topology from the true topology $\mathcal{G}$, we define the \emph{structural camouflage ratio (SCR)} as $\rho(\mathcal{H},\mathcal{G})=1-{|\mathcal{E}_{\mathcal H}\cap\mathcal{E}|}/{|\mathcal{E}_{\mathcal H}\cup\mathcal{E}|}$, where $\rho\in[0,1]$, with larger values indicating less structural overlap with $\mathcal{G}$. In particular, $\rho=0$ corresponds to identical topologies, whereas $\rho=1$ indicates no shared communication edges. Rather than relying on a single predetermined surrogate, \textsc{Mirage} further introduces a \emph{topology anonymity set}, inspired by the principle of $k$-anonymity~\citep{sweeney2002k}, to conceal $\mathcal{G}$ among multiple structurally feasible alternatives.

\begin{mdframed}[
  linecolor=black,
  linewidth=0.6pt,
  skipabove=4pt,   % ← 控制框“上面”的间距
  skipbelow=4pt,   % ← 控制框“下面”的间距
  innertopmargin=6pt,    % ← 框内顶部空白
  innerbottommargin=2pt  % ← 框内底部空白
]
\begin{definition}[Topology Anonymity Set]
Given a genuine topology $\mathcal{G}$ and a minimum structural deviation $\rho_0$, the topology anonymity set of $\mathcal{G}$ is defined as
$\mathcal{K}_{\rho_0}(\mathcal{G})=\{\mathcal{H}\in\Omega(\mathcal{G})\mid\rho(\mathcal{H},\mathcal{G})\geq\rho_0\}$. The corresponding anonymity level is characterized by $K_{\text{topo}}=|\mathcal{K}_{\rho_0}(\mathcal{G})|$, i.e., the number of feasible topologies satisfying the minimum structural deviation from the genuine
topology $\mathcal{G}$.
\end{definition}
\end{mdframed}
The topology anonymity set captures both structural deviation and candidate diversity: $\rho_0$ controls deviation from the genuine topology, while $K_{\text{topo}}$ reflects the number of feasible alternatives. In practice, \textsc{Mirage} constructs $\mathcal{K}_{\rho_0}(\mathcal{G})$ via order-preserving edge rewiring over $\mathcal{U}_{\pi}$ while preserving connectivity to the decision agent and respecting the original execution order. A candidate satisfying $\rho(\mathcal{H},\mathcal{G}) \ge \rho_0$ is selected as the phantom topology $\mathcal{G}'$, inducing $\mathcal{E}_{p}=\mathcal{E}'\setminus\mathcal{E}$ and $\mathcal{E}_{h}=\mathcal{E}\setminus\mathcal{E}'$ for phantom and concealed genuine relations, respectively.

\vspace{-0.5em}

\subsection{Semantic Edge Realization}\label{sec4.2}

\vspace{-0.5em}

Given the phantom topology $\mathcal{G}'$, the next challenge is to make its structural relations observable to the adversary through semantic dependencies. Since topology inference attacks exploit semantic dependencies among observable reasoning traces, merely constructing $\mathcal{G}'$ is insufficient: phantom edges must be manifested as plausible semantic dependencies, while genuine edges absent from $\mathcal{G}'$ should be obscured to suppress their source-target dependencies. \textsc{Mirage} therefore performs phantom structure injection in two directions: \ding{172} \text{phantom edge materialization} and \ding{173} \text{genuine edge obfuscation}. Both operations are applied to the adversary-facing context and do not alter the underlying communication topology used for task execution. More implementation details are provided in App.~C.

\begin{wrapfigure}{R}{0.53\textwidth}
\begin{minipage}{0.53\textwidth}
\vspace{-0.4in}
\begin{algorithm}[H]
    \caption{Overview of \textsc{Mirage}}
    \label{alg:mirage}
    \renewcommand{\algorithmicrequire}{\textbf{Input:}}
    \renewcommand{\algorithmicensure}{\textbf{Output:}}
    \begin{algorithmic}[1]
        \REQUIRE MAS $\mathcal{S}=(\mathcal{P},\mathcal{G})$, query $q$, deviation threshold $\rho_0$, generator $\mathcal{M}_{\phi}$.
        \ENSURE Protected response $\widetilde{\mathcal{S}}(q)$.

        \item[] \textcolor{cyan}{\textsc{// I. Phantom Topology Synthesis}}
        \STATE Construct $\mathcal{K}_{\rho_0}(\mathcal{G})$ via edge rewiring.
        \STATE Select $\mathcal{G}'=(\mathcal{A},\mathcal{E}')\in\mathcal{K}_{\rho_0}(\mathcal{G})$.
        \STATE $\mathcal{E}_{p}\gets\mathcal{E}'\setminus\mathcal{E},\quad \mathcal{E}_{h}\gets\mathcal{E}\setminus\mathcal{E}'$.

        \item[] \textcolor{cyan}{\textsc{// II. Semantic Edge Realization}}
        \FOR{each agent $a_i\in\mathcal{A}$ in execution order}
            \STATE $\mathcal{O}_i\hspace{-0.2em}\gets\hspace{-0.2em}\{r_j\hspace{-0.2em}\mid\hspace{-0.2em}(a_j,a_i)\hspace{-0.2em}\in\hspace{-0.2em}\mathcal{E}\}$; $r_i\hspace{-0.2em}\gets\hspace{-0.2em}\text{LLM}(p_i,q,\mathcal{O}_i)$.
            \STATE Materialize $s_{j\rightarrow i}^{\text{ph}}$ for $(a_j,a_i)\in\mathcal{E}_{p}$.
            \STATE Generate $\mathcal{R}_i=\{\hat r_i^{(m)}\}_{m=1}^{M}$ from $r_i$.
            \STATE Select $\widetilde r_i$ via dependency-aware obfuscation.
        \ENDFOR

        \item[] \textcolor{cyan}{\textsc{// III. Protected MAS Execution}}
        \STATE Retain $\{r_i\}$ as the execution view.
        \STATE Build the exposure view using $\{s_{j\rightarrow i}^{\text{ph}}\}$ and $\{\widetilde r_i\}$.
        \STATE Generate the protected response $\widetilde{\mathcal{S}}(q)$.
        \RETURN $\widetilde{\mathcal{S}}(q)$.
    \end{algorithmic}
\end{algorithm}
\vspace{-3em}
\end{minipage}
\end{wrapfigure}

\textbf{Phantom Edge Materialization.} 
For each phantom edge $(a_j,a_i)\in\mathcal{E}_p$, \textsc{Mirage} creates semantic evidence $s_{j\rightarrow i}^{\text{ph}}$ consistent with a dependency from agent $a_j$ to $a_i$. Specifically, $s_{j\rightarrow i}^{\text{ph}}$ is generated by an LLM-based generator $\mathcal{M}_{\phi}$ conditioned on the source agent's profile $p_j$ and execution output $r_j$, the target agent's profile $p_i$, the query $q$, and the target's current task-relevant contextual state $c_i$:
\begin{equation}
s_{j\rightarrow i}^{\text{ph}}
=
\mathcal{M}_{\phi}(p_j,r_j,p_i,q,c_i).
\end{equation}
Here, $c_i$ denotes the task-relevant context available to $a_i$. To ensure effective phantom edge materialization, the generated evidence follows three principles: \ding{172} \textit{role consistency}, aligning the content with the source agent's capabilities; \ding{173} \textit{task relevance}, providing useful information for the target agent; and \ding{174} \textit{dependency plausibility}, maintaining semantic consistency with $r_j$ while forming a plausible dependency toward $a_i$ without introducing unsupported information or topology cues.

\textbf{Genuine Edge Obfuscation.}
Phantom edge materialization alone cannot conceal genuine edges that are absent from $\mathcal{G}'$. For each agent $a_i$, \textsc{Mirage} therefore suppresses the observable dependency evidence directly associated with its hidden genuine predecessors $\mathcal{P}_{i}^{h}=\{a_j\mid(a_j,a_i)\in\mathcal{E}_{h}\}$, while preserving the task-relevant semantics of its original execution trace $r_i$. Specifically, \textsc{Mirage} adopts controlled paraphrasing~\citep{bandel2022quality} to generate a candidate set $\mathcal{R}_i=\{\hat r_i^{(m)}\}_{m=1}^{M}$ from $r_i$. The paraphrasing process varies the expression and organization of the original trace while preserving its task-relevant facts, reasoning, and conclusions. The protected trace is then selected according to
\begin{equation}
\widetilde{r}_i=\arg\min_{\hat r\in\mathcal{R}_i}\left[\mathcal{D}_{\text{sem}}(\hat r,r_i)+\lambda\sum_{a_j\in\mathcal{P}_{i}^{h}}\mathcal{S}_{\text{dep}}(r_j,\hat r)\right],
\label{eq:geo}
\end{equation}
where $\mathcal{D}_{\text{sem}}(\cdot,\cdot)$ measures semantic distortion, $\mathcal{S}_{\text{dep}}(\cdot,\cdot)$ measures semantic dependency, with higher values indicating stronger dependency, and $\lambda$ balances semantic preservation and dependency suppression. By suppressing genuine dependency cues while preserving task-relevant semantics, genuine edge obfuscation complements phantom edge materialization to shift the observable dependency structure away from $\mathcal{G}$ and toward $\mathcal{G}'$ without altering the underlying task execution.

\vspace{-0.5em}
\subsection{Protected MAS Execution}
\label{sec:dual_view}
\vspace{-0.5em}

Given $\mathcal{G}'$, \textsc{Mirage} decouples task execution from adversary-facing exposure to preserve utility while concealing the genuine topology. Specifically, the protected MAS operates through two views:
\vspace{-0.3em}
\begin{itemize}[leftmargin=*, itemindent=0em] 
\renewcommand{\labelitemi}{$\diamond$} 
\item \textit{Execution View.} Each agent executes over the genuine topology $\mathcal{G}$, receiving outputs only from its predecessors: $\mathcal{O}_i=\{r_j\mid(a_j,a_i)\in\mathcal{E}\}$ and $r_i=\text{LLM}(p_i,q,\mathcal{O}_i)$. The phantom topology does not participate in internal communication, and all genuine message passing remains governed by $\mathcal{G}$.

\item \textit{Exposure View.} Before exposure, \textsc{Mirage} obfuscates genuine dependencies in $\mathcal{E}_h$ to obtain $\widetilde r_i$ and incorporates phantom evidence from $\mathcal{E}_p$ to construct $r_i^{\text{exp}}$. The resulting semantic dependencies are shaped away from $\mathcal{G}$ and toward $\mathcal{G}'$.
\end{itemize}
\vspace{-0.3em}

This dual-view design retains $\mathcal{G}$ for task execution while exposing dependency evidence aligned with $\mathcal{G}'$, thereby concealing the genuine communication topology with minimal impact on task utility.

\begin{figure*}[t]
    \centering
    \begin{minipage}{\textwidth}
        \centering
        \subfigure[\normalsize MMLU]{\includegraphics[width=0.24\textwidth]{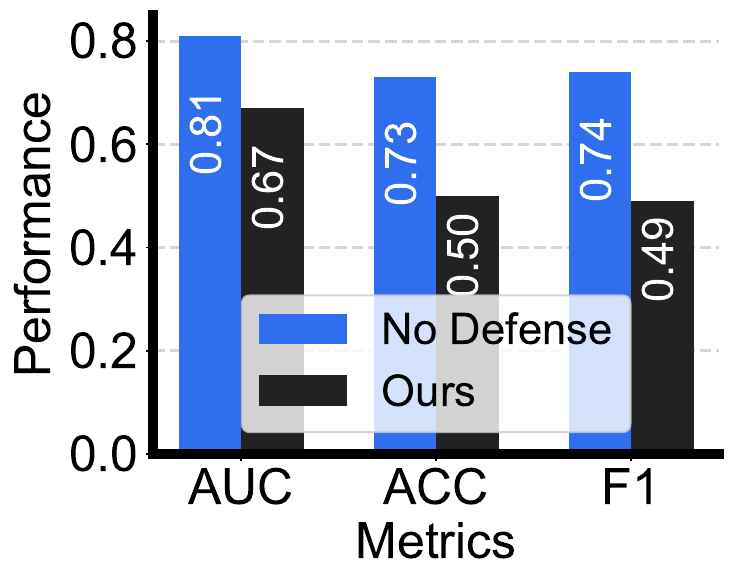}}
        \subfigure[\normalsize GSM8K]{\includegraphics[width=0.24\textwidth]{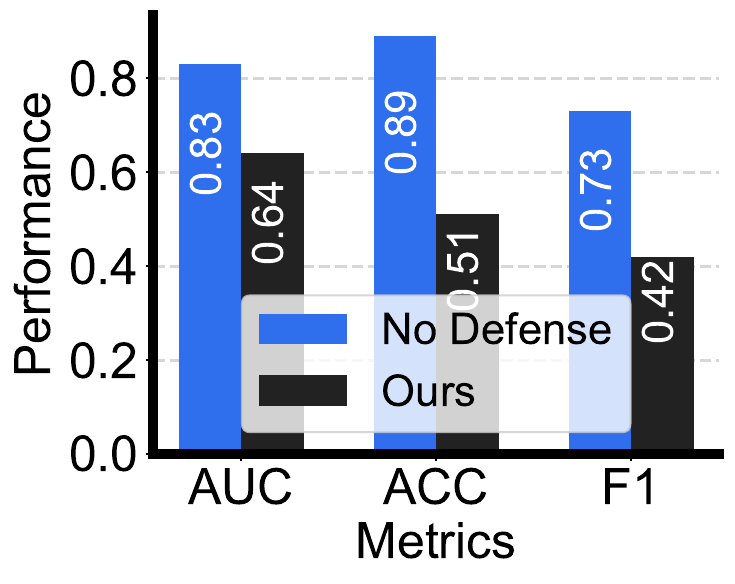}}
        \subfigure[\normalsize SVAMP]{\includegraphics[width=0.24\textwidth]{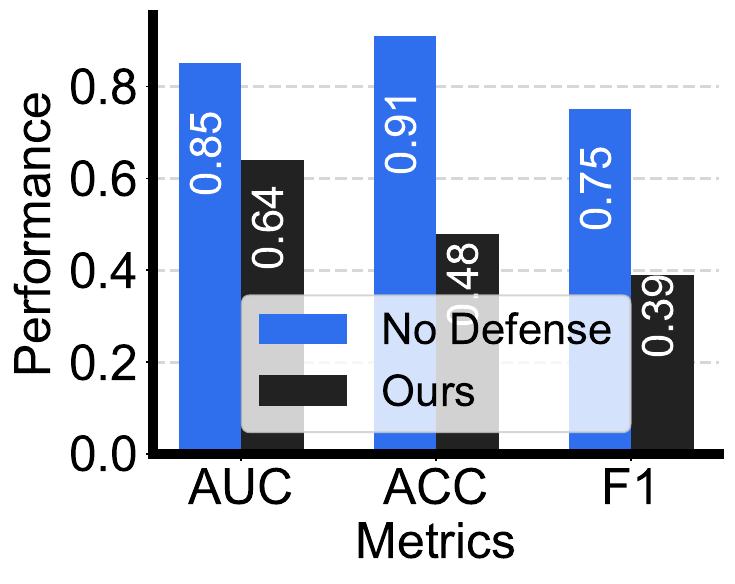}}
        \subfigure[\normalsize HumanEval]{\includegraphics[width=0.24\textwidth]{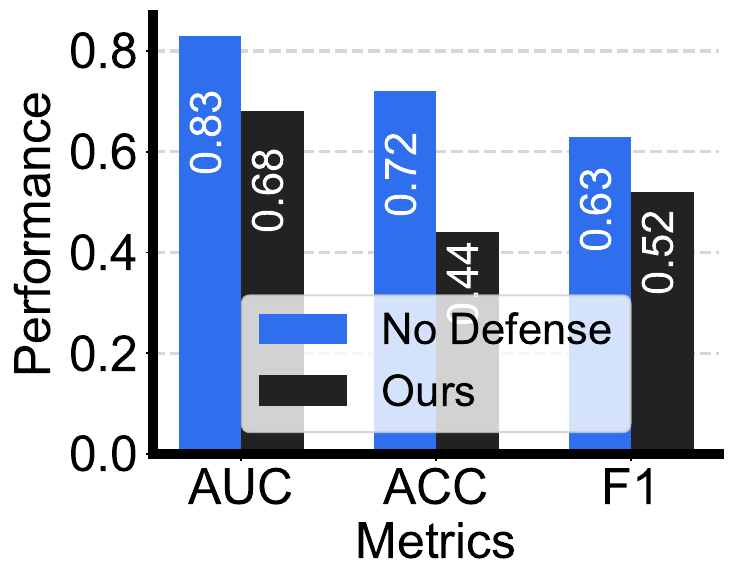}}
     \end{minipage}\\ \vspace{-1em}
   \caption{Comparison of topology inference performance using G-Designer before defense (\textsf{No Defense}) and with \textsc{Mirage} (\textsf{Ours}) across four benchmark datasets in terms of AUC, ACC, and F1.}
    \label{fig:exp3}
    \vspace{-1em}
\end{figure*}

\begin{figure*}[t]
    \centering
    \begin{minipage}{\textwidth}
        \centering
        \subfigure[\normalsize MMLU]{\includegraphics[width=0.24\textwidth]{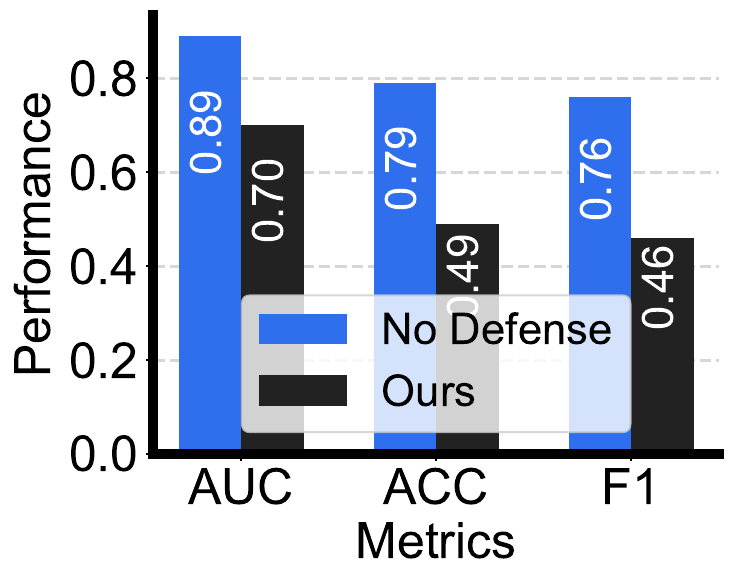}}
        \subfigure[\normalsize GSM8K]{\includegraphics[width=0.24\textwidth]{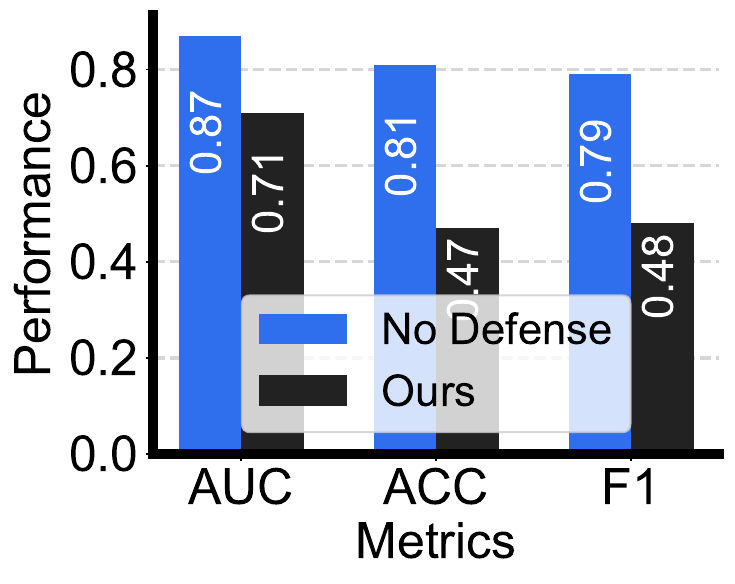}}
        \subfigure[\normalsize SVAMP]{\includegraphics[width=0.24\textwidth]{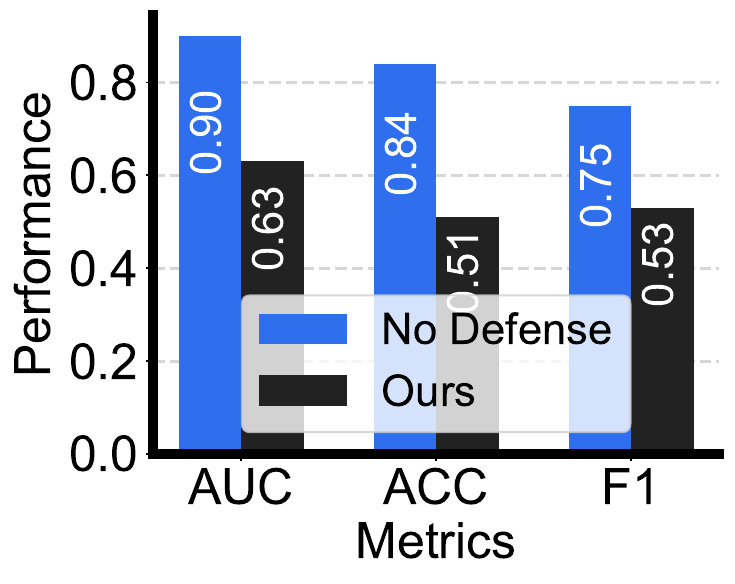}}
        \subfigure[\normalsize HumanEval]{\includegraphics[width=0.24\textwidth]{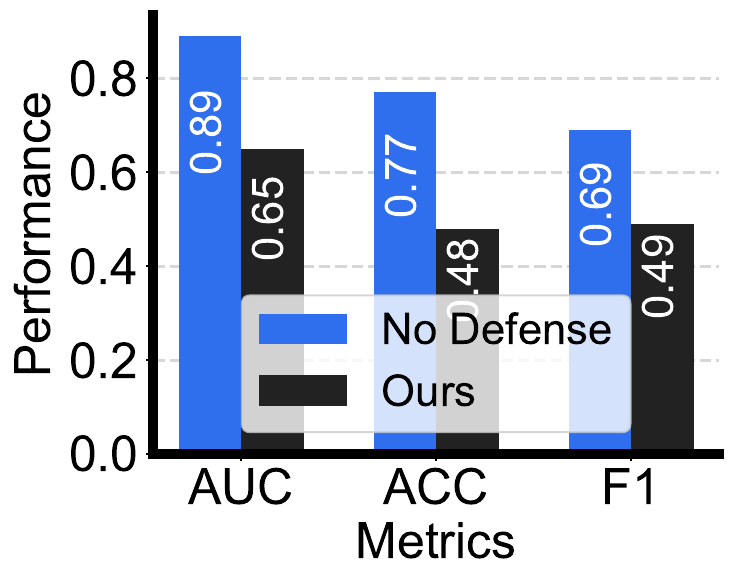}}
     \end{minipage}\\ \vspace{-1em}
     \caption{Comparison of topology inference performance using AGP before defense (\textsf{No Defense}) and with \textsc{Mirage} (\textsf{Ours}) across four benchmark datasets in terms of AUC, ACC, and F1.}
    \label{fig:exp4}
    \vspace{-1em}
\end{figure*}

\begin{figure*}[h]
    \centering
    \begin{minipage}{\textwidth}
        \centering
        \subfigure[\normalsize MMLU]{\includegraphics[width=0.24\textwidth]{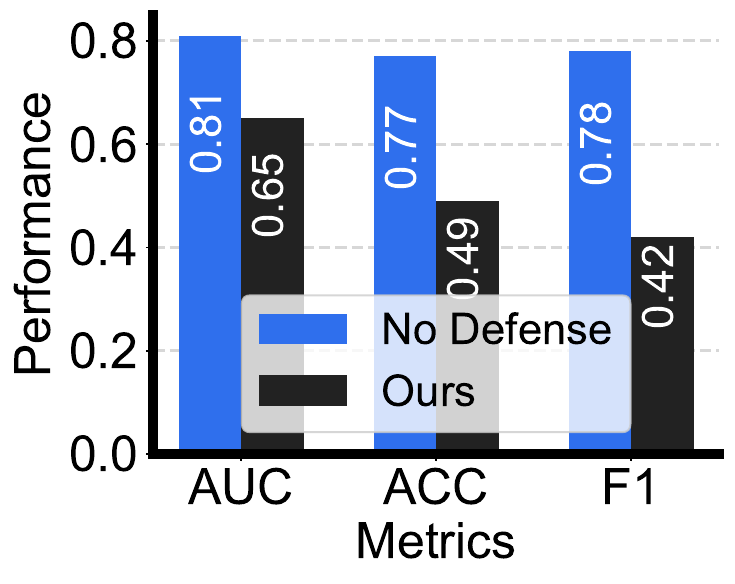}}
        \subfigure[\normalsize GSM8K]{\includegraphics[width=0.24\textwidth]{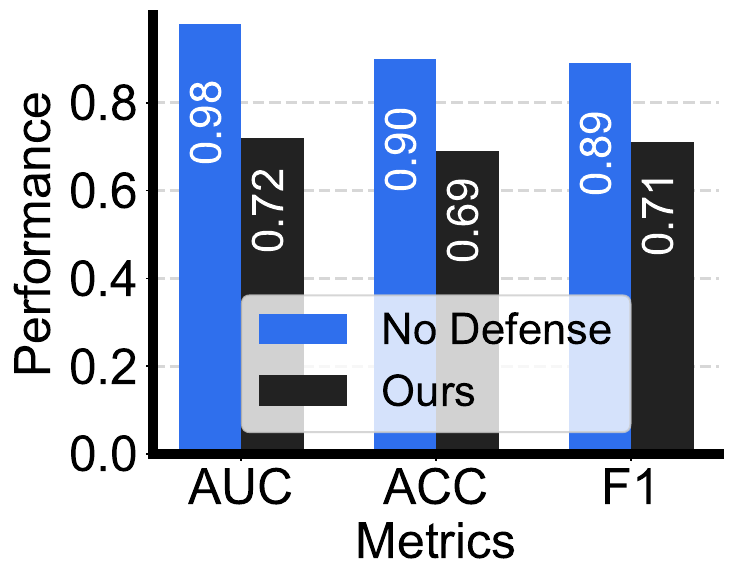}}
        \subfigure[\normalsize SVAMP]{\includegraphics[width=0.24\textwidth]{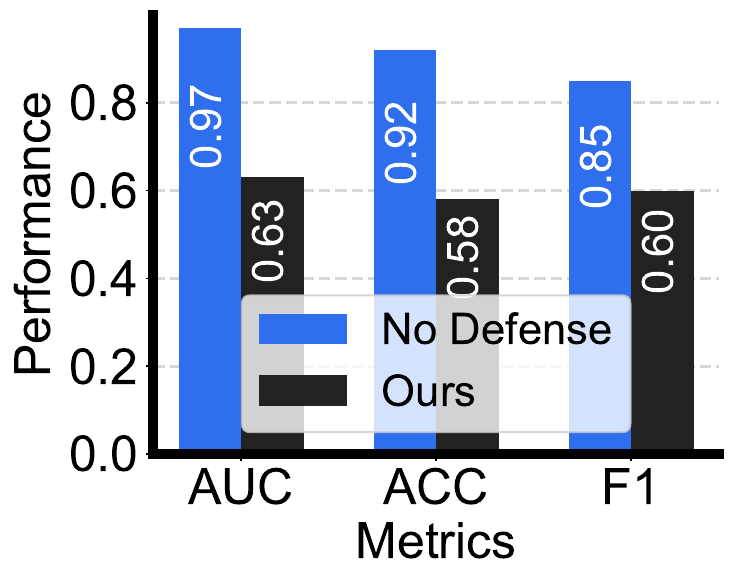}}
        \subfigure[\normalsize HumanEval]{\includegraphics[width=0.24\textwidth]{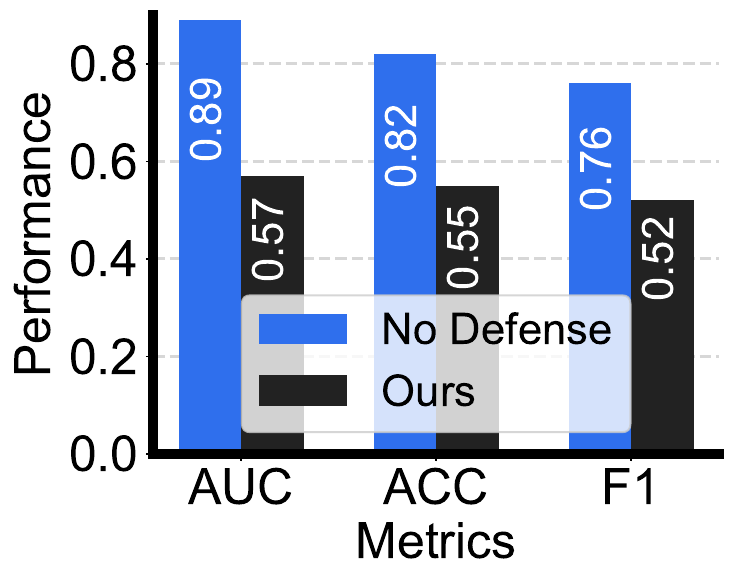}}
     \end{minipage}\\ \vspace{-1em}
     \caption{Comparison of topology inference performance using ARG-Designer before defense (\textsf{No Defense}) and with \textsc{Mirage} (\textsf{Ours}) across four benchmark datasets in terms of AUC, ACC, and F1.}
    \label{fig:exp5}
    \vspace{-0.5em}
\end{figure*}

\begin{table*}[t]
\centering
\caption{Comparison of topology inference performance under different defense methods using G-Designer across four benchmark datasets in terms of AUC, ACC, and F1.}
\vspace{-0.5em}
\label{tab:tab1}
\resizebox{\textwidth}{!}{
\begin{tabular}{l|ccc|ccc|ccc|ccc}
\toprule
\multirow{2}{*}{\textbf{Method}} &
\multicolumn{3}{c|}{\texttt{MMLU}} &
\multicolumn{3}{c|}{\texttt{GSM8K}} &
\multicolumn{3}{c|}{\texttt{SVAMP}} &
\multicolumn{3}{c}{\texttt{HumanEval}} \\
\cmidrule(lr){2-4}
\cmidrule(lr){5-7}
\cmidrule(lr){8-10}
\cmidrule(lr){11-13}
&
\textbf{AUC} & \textbf{ACC} & \textbf{F1} &
\textbf{AUC} & \textbf{ACC} & \textbf{F1} &
\textbf{AUC} & \textbf{ACC} & \textbf{F1} &
\textbf{AUC} & \textbf{ACC} & \textbf{F1} \\
\midrule

 \textsc{Instruction}
& 0.79 & 0.59 & 0.62
& 0.81 & 0.73 & 0.64
& 0.72 & 0.74 & 0.69
& 0.78 & 0.62 & 0.59 \\

 \textsc{Delimiters}
& 0.76 & 0.61 & 0.64
& 0.80 & 0.76 & 0.67
& 0.70 & 0.68 & 0.71
& 0.76 & 0.65 & 0.56 \\

 \textbf{\textsc{Mirage} (Ours)}
& \cellcolor{cyan!15}\textbf{0.67} 
& \cellcolor{cyan!15}\textbf{0.50} 
& \cellcolor{cyan!15}\textbf{0.49}
& \cellcolor{cyan!15}\textbf{0.64} 
& \cellcolor{cyan!15}\textbf{0.51} 
& \cellcolor{cyan!15}\textbf{0.42}
& \cellcolor{cyan!15}\textbf{0.64} 
& \cellcolor{cyan!15}\textbf{0.48} 
& \cellcolor{cyan!15}\textbf{0.39}
& \cellcolor{cyan!15}\textbf{0.68} 
& \cellcolor{cyan!15}\textbf{0.44} 
& \cellcolor{cyan!15}\textbf{0.52} \\

\bottomrule
\end{tabular}
}
\vspace{-1em}
\end{table*}

\begin{figure*}[t]
    \centering
    \begin{minipage}{\textwidth}
        \centering
        \subfigure[\normalsize MMLU]{\includegraphics[width=0.24\textwidth]{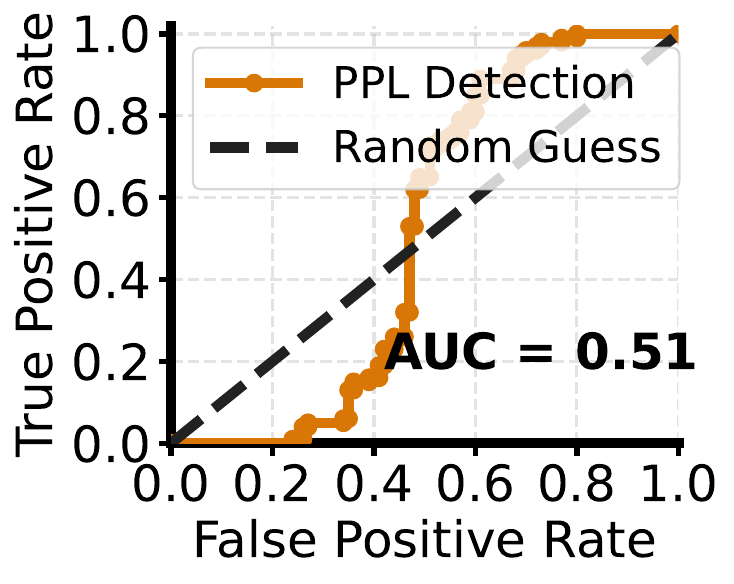}}
        \subfigure[\normalsize GSM8K]{\includegraphics[width=0.24\textwidth]{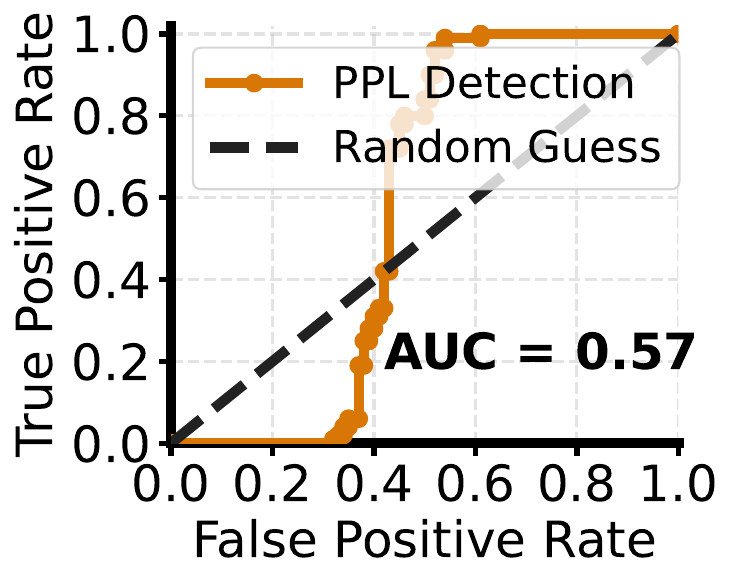}}
        \subfigure[\normalsize SVAMP]{\includegraphics[width=0.24\textwidth]{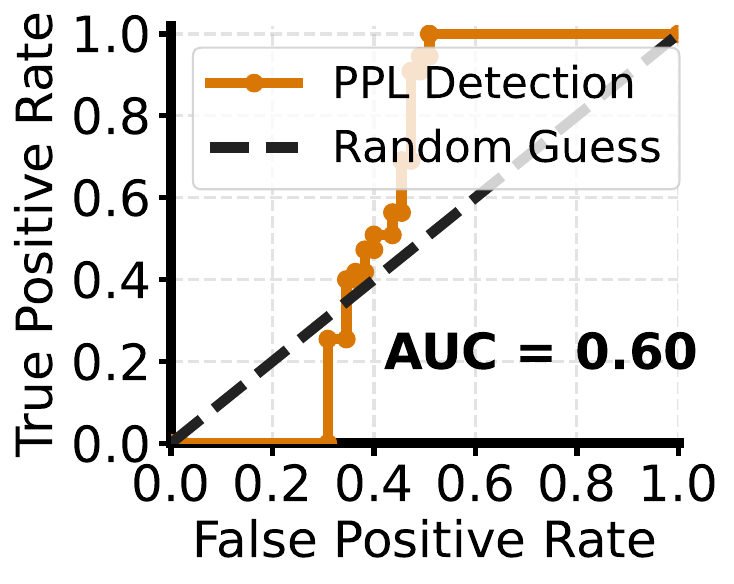}}
        \subfigure[\normalsize HumanEval]{\includegraphics[width=0.24\textwidth]{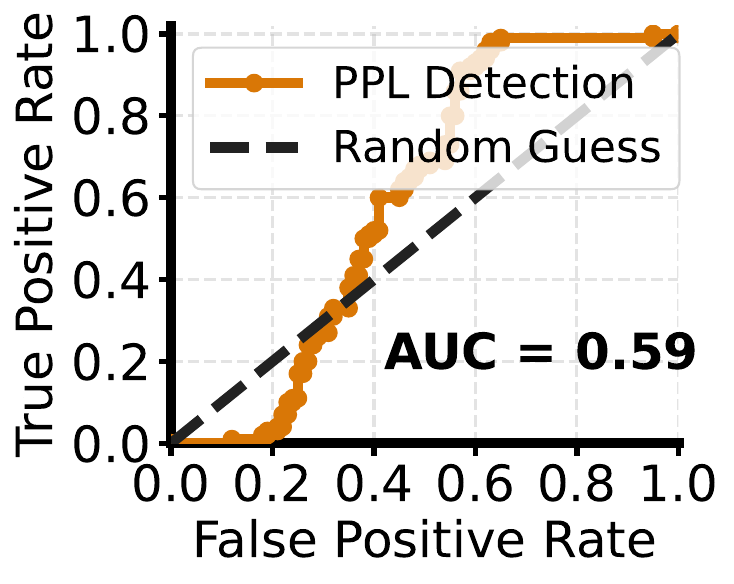}}
     \end{minipage}\\ \vspace{-1em}
   \caption{Overall ROC curves of PPL-based attack detection across four benchmark datasets, with the corresponding detection AUC reported and random guessing shown as a reference.}
    \label{fig:exp6}
    \vspace{-1em}
\end{figure*}

\begin{figure*}[h]
    \centering
    \begin{minipage}{\textwidth}
        \centering
        \subfigure[\normalsize MMLU]{\includegraphics[width=0.24\textwidth]{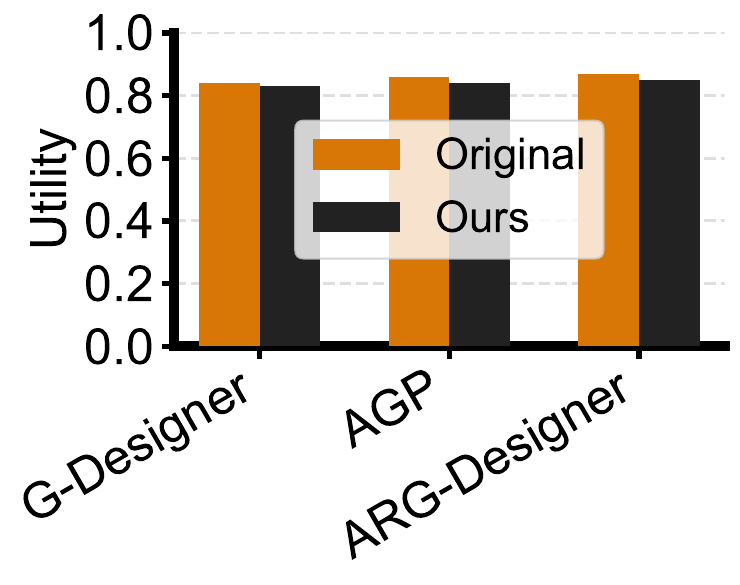}}
        \subfigure[\normalsize GSM8K]{\includegraphics[width=0.24\textwidth]{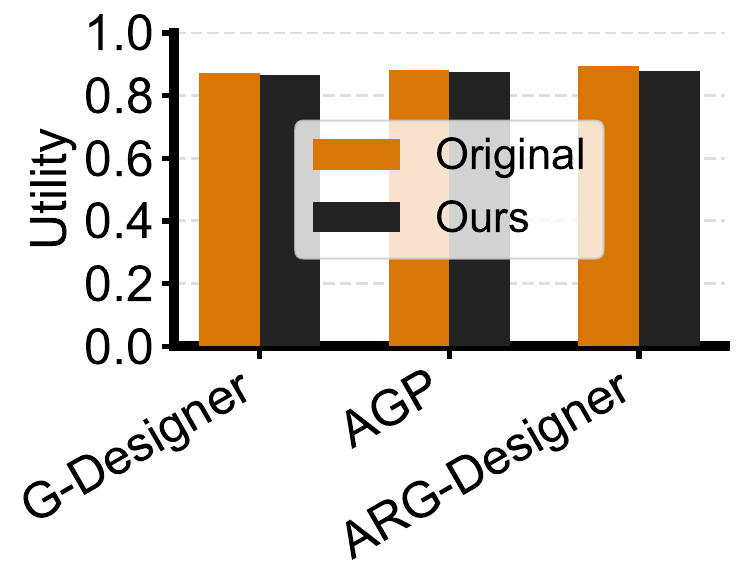}}
        \subfigure[\normalsize SVAMP]{\includegraphics[width=0.24\textwidth]{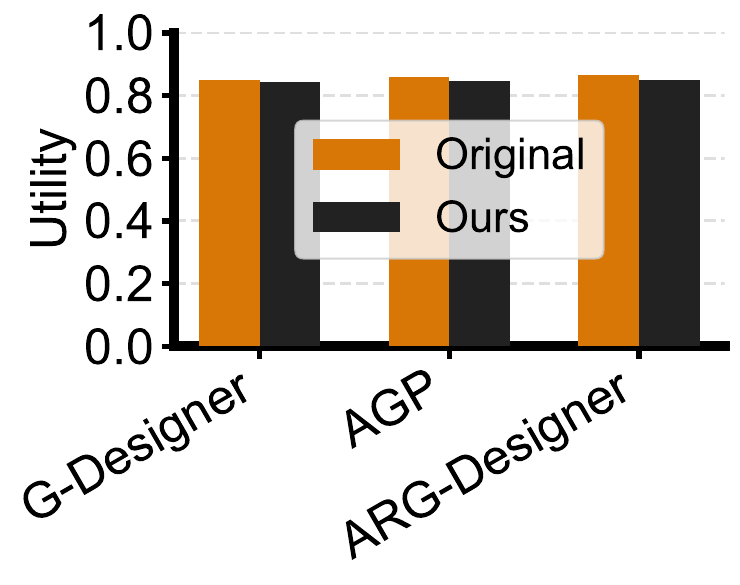}}
        \subfigure[\normalsize HumanEval]{\includegraphics[width=0.24\textwidth]{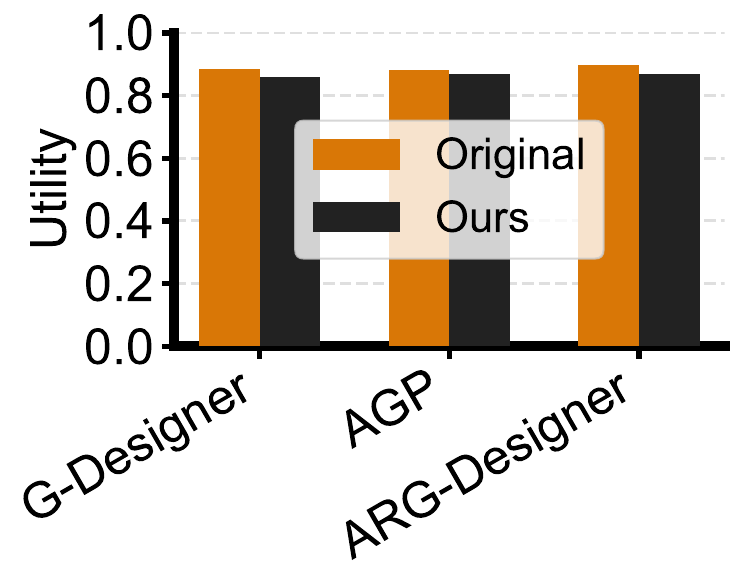}}
     \end{minipage}\\ \vspace{-1em}
    \caption{Comparison of task utility before defense (\textsf{Original}) and with \textsc{Mirage} (\textsf{Ours}) across three topology optimization strategies and four benchmark datasets, measured by task accuracy.
}
    \label{fig:exp7}
    \vspace{-1em}
\end{figure*}

\begin{figure*}[h]
    \centering
    \begin{minipage}{\textwidth}
        \centering
        \subfigure[\normalsize Ablation Study]{\includegraphics[width=0.24\textwidth]{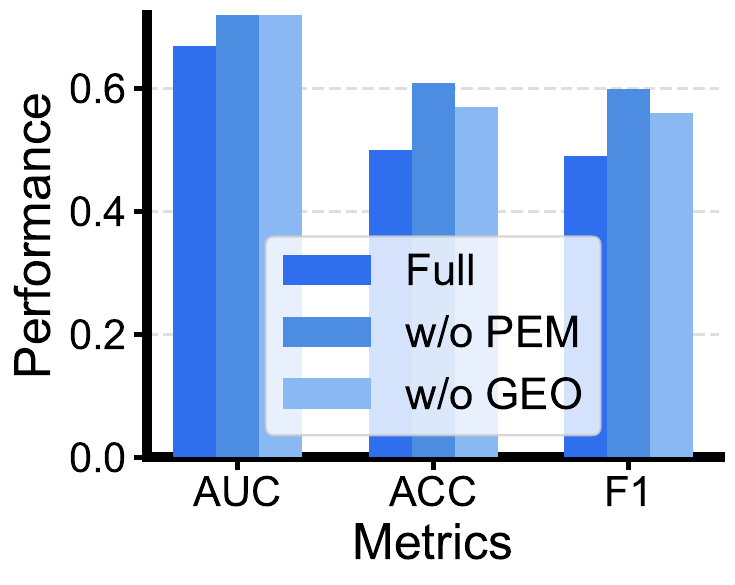}}
        \subfigure[\normalsize Parameter $\rho_0$]{\includegraphics[width=0.24\textwidth]{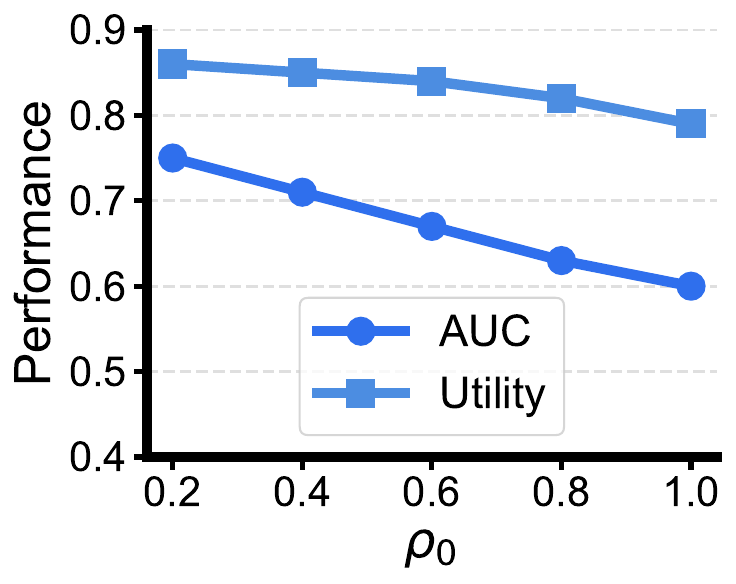}}
        \subfigure[\normalsize Parameter $M$]{\includegraphics[width=0.24\textwidth]{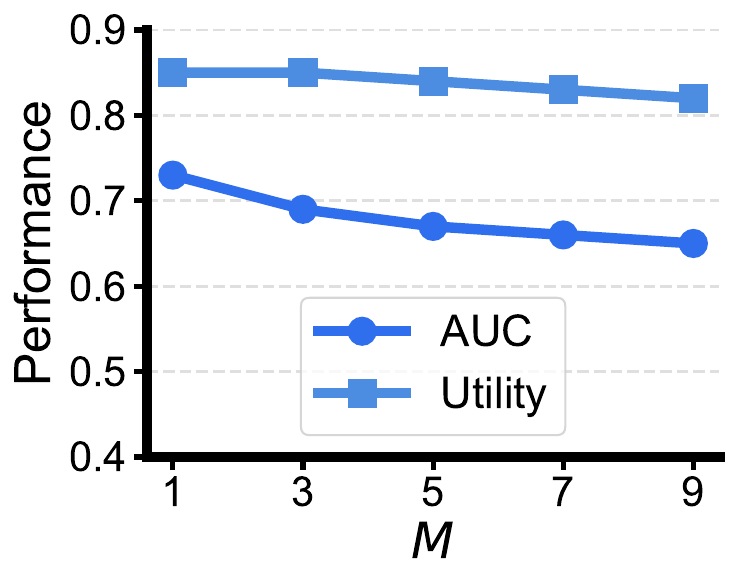}}
        \subfigure[\normalsize Parameter $\lambda$]{\includegraphics[width=0.24\textwidth]{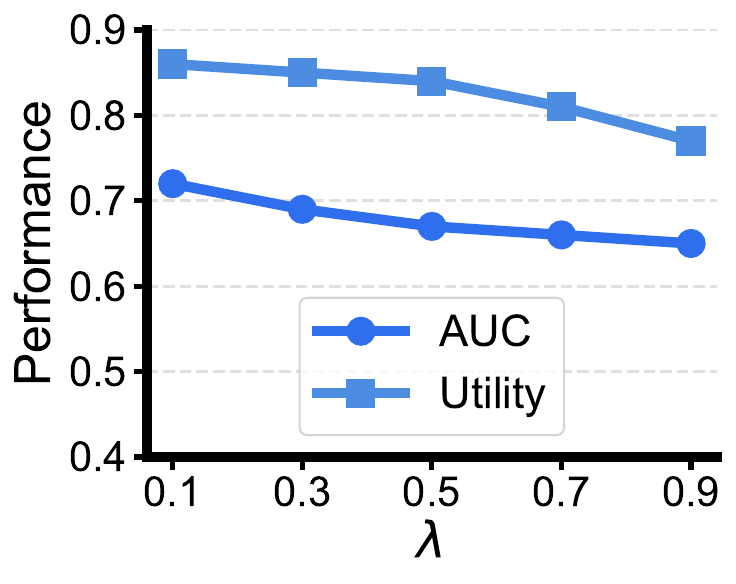}}
     \end{minipage}\\ \vspace{-1em}
    \caption{Ablation and parameter analyses of \textsc{Mirage} under different experimental settings. (a) Effects of removing key defense components. (b)--(d) Effects of the structural deviation threshold $\rho_0$, number of paraphrase candidates $M$, and dependency suppression weight $\lambda$, respectively.}
    \label{fig:exp8}
    \vspace{-0.5em}
\end{figure*}

\section{Experiments}\label{sec5}
\vspace{-0.5em}
In this section, we conduct extensive experiments to evaluate \textsc{Mirage} in protecting the communication topology of LLM-based MAS while preserving task utility across diverse tasks and communication topology configurations. Specifically, we investigate the following research questions:

\vspace{-0.5em}
\begin{itemize}[leftmargin=*, itemindent=0em]
\renewcommand{\labelitemi}{$\diamond$}
\item \textbf{RQ1:} How effectively does \textsc{Mirage} defend against communication topology inference attacks?

\item \textbf{RQ2:} How well does \textsc{Mirage} preserve task utility while protecting communication topologies?

\item \textbf{RQ3:} How do the components and parameter settings of \textsc{Mirage} affect its defense effectiveness?

\item \textbf{RQ4:} How does \textsc{Mirage} conceal the genuine communication topology in representative cases?
\end{itemize}

\vspace{-0.5em}
\subsection{Experimental Settings}\label{sec5.1}
\vspace{-0.5em}
\textbf{Datasets.}
We evaluate \textsc{Mirage} on four datasets covering three representative task domains. For \textit{general reasoning}, we use \texttt{MMLU}~\citep{DBLP:conf/iclr/HendrycksBBZMSS21}, which evaluates knowledge and reasoning capabilities across diverse subject areas. For \textit{mathematical reasoning}, we adopt \texttt{GSM8K}~\citep{cobbe2021training} and \texttt{SVAMP}~\citep{patel2021nlp}, both of which require multi-step reasoning to solve mathematical word problems. For \textit{code generation}, we use \texttt{HumanEval}~\citep{chen2021evaluating}, which evaluates the functional correctness of generated programs. Consistent with prior work~\citep{wu2026cia}, we sample 100 tasks from each dataset for evaluation. More details are provided in App.~D.

\textbf{MAS Frameworks.}
Three representative topology optimization frameworks are adopted to construct the target MAS: \textit{G-Designer}~\citep{zhang2025g}, \textit{AGP}~\citep{DBLP:conf/ecai/LiZLW25}, and \textit{ARG-Designer}~\citep{li2026assemble}. These frameworks employ different topology optimization strategies, enabling evaluation of \textsc{Mirage} across diverse communication structures.
 More details are provided in App.~D.

\textbf{Baselines.} We first compare \textsc{Mirage} against the original MAS without any protection (\textit{No Defense}). For a more comprehensive evaluation, we further consider three potential defenses commonly used against prompt injection attacks~\citep{liu2024formalizing,zhan2025adaptive}: \ding{172} \textit{Instructional Prevention} (\textit{Instruction}), \ding{173} \textit{Delimiters}, and \ding{174} \textit{PPL Detection}. These methods respectively represent instruction-level prevention, input isolation, and detection-based defense. More details are provided in App.~D.

\textbf{Metrics \& Parameters.}
We evaluate \textsc{Mirage} in terms of topology protection and task utility. Following prior work~\citep{wu2026cia}, topology inference is measured by AUC, ACC, and F1 over candidate communication edges, where AUC closer to 0.5 and lower ACC/F1 indicate stronger protection. Task utility is measured by accuracy across the evaluated benchmark tasks. We select $\rho_0$, $M$, and $\lambda$ from $\{0.2,0.4,0.6,0.8,1.0\}$, $\{1,3,5,7,9\}$, and $\{0.1,0.3,0.5,0.7,0.9\}$, respectively. More detailed implementation and parameter settings are provided in App.~D.

\vspace{-1em}
\subsection{Results \& Discussion}\label{sec5.2}
\vspace{-0.5em}

\begin{wraptable}{r}{0.65\textwidth}
\vspace{-13pt}
\setlength\tabcolsep{4.5pt}
\small
\centering
\caption{Statistics of communication topologies generated by three topology optimization methods across four datasets. $\bar{N}$ and $\bar{E}$ denote the average numbers of agents and edges, respectively.}
\label{tab:tab2}
\vspace{-0.5em}
\resizebox{\linewidth}{!}{
\begin{tabular}{lcccccccc}
\toprule
\multirow{2}{*}{\textbf{Method}}
& \multicolumn{2}{c}{\texttt{MMLU}}
& \multicolumn{2}{c}{\texttt{GSM8K}}
& \multicolumn{2}{c}{\texttt{SVAMP}}
& \multicolumn{2}{c}{\texttt{HumanEval}} \\
\cmidrule(lr){2-3}
\cmidrule(lr){4-5}
\cmidrule(lr){6-7}
\cmidrule(lr){8-9}
& $\bar{N}$ & $\bar{E}$
& $\bar{N}$ & $\bar{E}$
& $\bar{N}$ & $\bar{E}$
& $\bar{N}$ & $\bar{E}$ \\
\midrule
G-Designer
& 7.00 & 8.99
& 5.00 & 8.19
& 5.00 & 8.15
& 6.00 & 11.38 \\

AGP
& 6.00 & 10.87
& 5.00 & 8.45
& 5.00 & 8.41
& 6.00 & 11.54 \\

ARG-Designer
& 5.42 & 7.84
& 3.07 & 3.14
& 3.05 & 3.10
& 4.24 & 5.49 \\
\bottomrule
\end{tabular}}
\vspace{-10pt}
\end{wraptable}

\textbf{Defense Effectiveness ($\triangleright$ RQ1)}.
Figs.~\ref{fig:exp3}--\ref{fig:exp5} comprehensively compare topology inference performance before (\textit{No Defense}) and after applying \textsc{Mirage} (\textit{Ours}) across three representative topology optimization strategies and four benchmark datasets. Our \textsc{Mirage} consistently reduces AUC, ACC, and F1, demonstrating its effectiveness in concealing the genuine communication topology. We further compare \textsc{Mirage} with existing defenses. As shown in Table~\ref{tab:tab1}, \textsc{Mirage} achieves consistently lower topology inference performance than \textsc{Instruction} and \textsc{Delimiters}. We additionally evaluate PPL-based detection. Fig.~\ref{fig:exp6} reports its ROC curves and detection AUC, which measures the ability to distinguish topology-inference queries from benign inputs. Finally, Table~\ref{tab:tab2} summarizes the topology statistics across different settings. Despite substantial variations in structural complexity, the results in Figs.~\ref{fig:exp3}--\ref{fig:exp5} show that \textsc{Mirage} remains effective across all three topology optimization strategies. Overall, these results demonstrate the effectiveness and robustness of \textsc{Mirage} against topology inference attacks.

\textbf{Utility Preservation ($\triangleright$ RQ2)}. Fig.~\ref{fig:exp7} compares the task utility before and after applying \textsc{Mirage} across three topology optimization strategies and four benchmark datasets. Despite substantially reducing topology inference performance, \textsc{Mirage} largely preserves the original task utility across all evaluated settings, with only minor performance variations. This demonstrates that \textsc{Mirage} effectively conceals the underlying communication topology while introducing only minor task-utility degradation, achieving a favorable balance between topology protection and task utility.

\textbf{Ablation Study ($\triangleright$ RQ3)}. We conduct an ablation study to systematically evaluate the contribution of the key components in \textsc{Mirage} using G-Designer on \texttt{MMLU}. As shown in Fig.~\ref{fig:exp8}(a), removing either phantom edge materialization (\textit{w/o PEM}) or genuine edge obfuscation (\textit{w/o GEO}) consistently increases AUC, ACC, and F1, indicating substantially degraded topology concealment performance and demonstrating that both components are essential for effective topology concealment.

\textbf{Parameter Analysis ($\triangleright$ RQ3)}. We investigate the sensitivity of \textsc{Mirage} to three key parameters, $\rho_0$, $M$, and $\lambda$, using G-Designer on \texttt{MMLU}. As shown in Figs.~\ref{fig:exp8}(b)--(d), $\rho_0$ exhibits a clear trade-off between topology concealment and task utility, while the benefit of increasing $M$ gradually saturates beyond $M=5$. For $\lambda$, moderate values provide a better balance between dependency suppression and semantic preservation. Overall, these results demonstrate that appropriate parameter settings enable \textsc{Mirage} to effectively conceal the communication topology while preserving task utility.

\begin{figure}[t]
  \centering
\includegraphics[width=\linewidth]{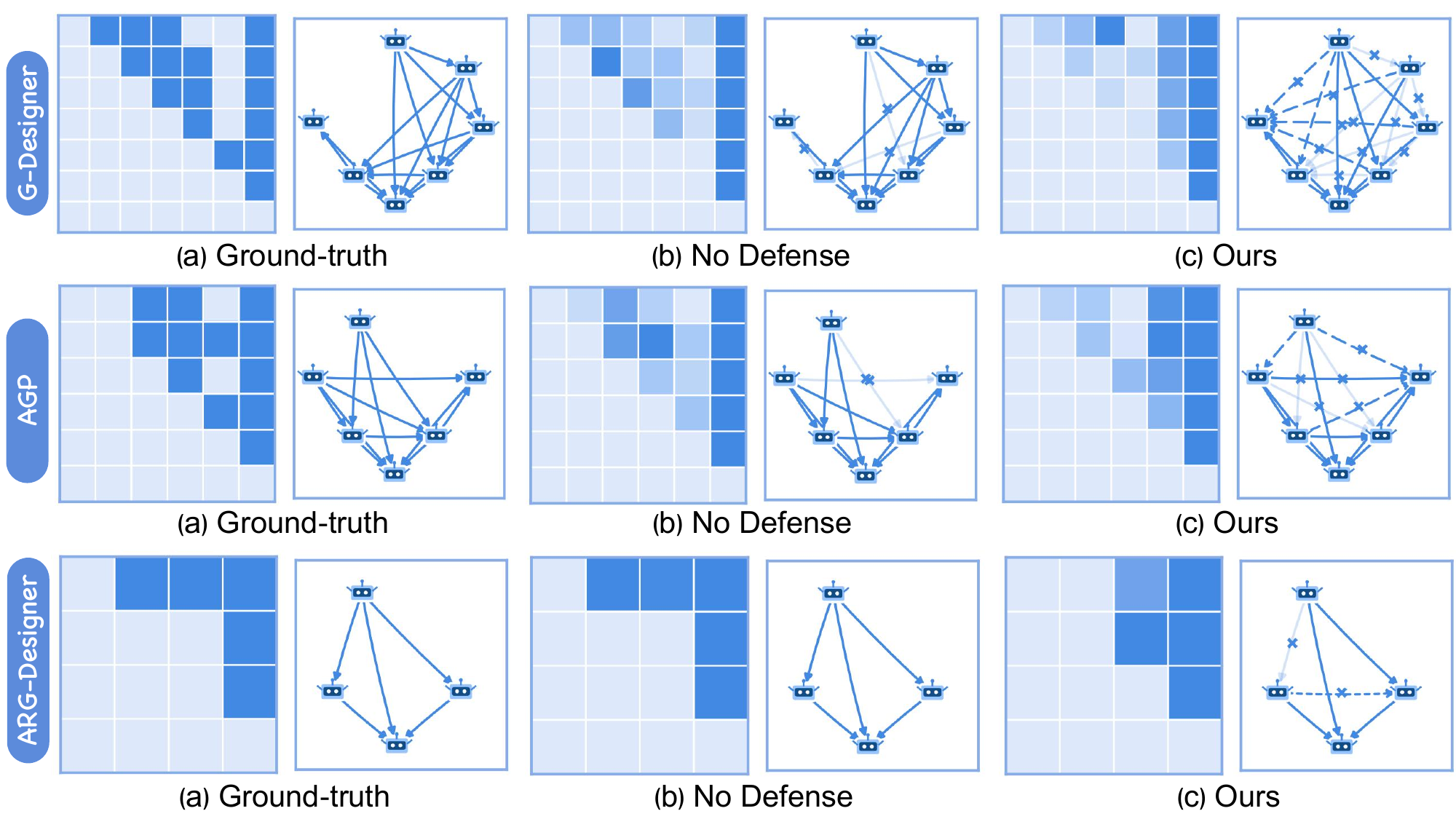}
\vspace{-1.7em}
  \caption{Visualization of communication topologies under G-Designer, AGP, and ARG-Designer, comparing the ground-truth topology (\textsf{Ground-truth}) with those inferred by CIA before defense (\textsf{No Defense}) and with \textsc{Mirage} (\textsf{Ours}) across diverse topology structures.}
\label{fig:exp9}
\vspace{-1.3em}
\end{figure}

\textbf{Case Study ($\triangleright$ RQ4).}
To intuitively demonstrate the topology concealment of \textsc{Mirage}, Fig.~\ref{fig:exp9} visualizes representative communication topologies generated by G-Designer, AGP, and ARG-Designer, together with the corresponding topologies inferred by CIA~\citep{wu2026cia} before and after defense. Without protection, CIA recovers most genuine communication edges, producing inferred structures highly similar to the ground-truth topologies. In contrast, after applying \textsc{Mirage}, the inferred topologies substantially deviate from the ground truth across all three topology optimization methods, with many genuine edges concealed or replaced by phantom relations. These cases visually confirm that \textsc{Mirage} substantially hinders the adversary from recovering the genuine communication structure.

\vspace{-0.5em}

\section{Related Work}\label{sec6}

\vspace{-0.5em}

This section reviews three lines of research closely related to our work: \ding{172} LLM-based multi-agent system, \ding{173} adversarial attacks against MAS, and \ding{174} privacy and security protection for MAS.

\textbf{LLM-based Multi-Agent Systems}. LLM-based multi-agent systems (MAS) coordinate multiple specialized agents through structured communication to solve complex tasks, demonstrating strong capabilities in software engineering~\citep{he2025llm,islam2024mapcoder,oueslati2026refagent}, scientific discovery~\citep{ghareeb2026multi,ghafarollahi2025sciagents}, and mathematical reasoning~\citep{lei2024macm,zhang2025debate4math}. Early systems typically adopt handcrafted communication structures, such as the sequential workflow in ChatDev~\citep{qian2024chatdev} and role-based collaboration in CAMEL~\citep{li2023camel} and MetaGPT~\citep{hong2024metagpt}. Recent studies further explore automated topology optimization to construct task-adaptive communication structures. Representative methods include G-Designer~\citep{zhang2025g}, AGP~\citep{DBLP:conf/ecai/LiZLW25}, and ARG-Designer~\citep{li2026assemble}, which optimize agent connectivity through graph-based generation or pruning. As these optimized topologies increasingly encode substantial computational investment and system design knowledge, protecting them from unauthorized inference becomes an important yet underexplored problem.

\textbf{Adversarial Attacks against MAS}. The growing adoption of LLM-based MAS has raised increasing concerns about their vulnerability to adversarial attacks~\citep{yu2025survey}. Existing studies primarily target agent behaviors and communication content, including prompt-based attacks~\citep{lee2025prompt,shahroz2025agents,arif2026conjunctive}, communication attacks~\citep{he2025red,yan2026attack}, and task disruption~\citep{amayuelas2024multiagent}, which inject malicious instructions or manipulate inter-agent interactions to compromise system behavior. Beyond content-level threats, CIA~\citep{wu2026cia} reveals a distinct risk to topology confidentiality by reconstructing MAS communication topologies from semantic dependencies under black-box access. Despite the resulting security and intellectual property risks, defenses against topology inference remain largely unexplored. This work fills this gap by introducing \textsc{Mirage} to protect MAS communication topologies against black-box inference.

\textbf{Privacy and Security Protection for MAS}. Existing defenses for MAS mainly focus on content-level security~\citep{zhou2026guardian,zhang2024psysafe,raza2026trism,miao2025blindguard,liu2024formalizing,zhan2025adaptive}. Representative approaches protect inter-agent collaboration through attack detection, trust management, or communication safeguards, such as BlindGuard~\citep{miao2025blindguard}, TRiSM~\citep{raza2026trism}, and GUARDIAN~\citep{zhou2026guardian}. However, these defenses protect message content or system behavior rather than the communication topology itself. Consequently, the confidentiality of MAS topologies remains largely overlooked. \textsc{Mirage} addresses this gap by concealing the genuine topology from black-box inference while preserving it for task execution.

\vspace{-0.5em}

\section{Conclusion}\label{sec7}
\vspace{-0.7em}

This paper investigates the emerging threat of communication topology inference in LLM-based multi-agent systems. To mitigate this threat, we propose \textsc{Mirage}, a topology-concealment framework that decouples genuine task execution from adversary-facing exposure. \textsc{Mirage} consists of three stages: \ding{172} phantom topology synthesis, \ding{173} semantic edge realization, and \ding{174} protected MAS execution. Specifically, it constructs structurally feasible phantom topologies and reshapes observable semantic dependencies through phantom edge materialization and genuine edge obfuscation, while preserving the genuine topology for task execution. Extensive experiments across multiple topology optimization methods and benchmark datasets demonstrate that \textsc{Mirage} substantially reduces topology inference effectiveness while largely preserving task utility. Overall, \textsc{Mirage} provides a practical and effective approach to protecting confidential MAS topologies against black-box inference attacks while largely preserving task utility. Further discussion of limitations and future directions is provided in App.~E.

\bibliography{main}
\bibliographystyle{iclr2027_conference}

\end{document}